\documentclass[]{amsart}

\usepackage[french]{babel}
\usepackage{amsmath}
\usepackage{amsthm}
\usepackage{amssymb}
\usepackage{amscd}
\usepackage{amsfonts}
\usepackage{amsbsy}
\usepackage{epsfig,afterpage}
\usepackage{psfrag}

\input cyracc.def
\font\tencyr=wncyr10
\def\cyr{\tencyr\cyracc}

\begin{document}

\title[Self-excited oscillations: from Poincar\'{e} to Andronov]{Self-excited oscillations:\\ from Poincar\'{e} to Andronov}
          
\author[J.M. Ginoux]{Jean-Marc Ginoux$^1$}

\address{$^1$ Laboratoire {\sc Protee}, I.U.T. de Toulon, Universit\'{e} du Sud, BP 20132, F-83957 La Garde Cedex, France, ginoux@univ-tln.fr, http://ginoux.univ-tln.fr}
\email{ginoux@univ-tln.fr}

\maketitle

\begin{abstract}

Till recently, the young Russian mathematician Aleksandr' Andronov was
considered by many scientists as the first to have applied the concept of
\textit{limit cycle}, introduced almost half a century before by Henri Poincar\'{e}, in order to
state the existence of self-sustained oscillations.

Consequently, if the discovery of a series of ``forgotten lectures'' given
by Poincar\'{e} at the \'{E}cole Sup\'{e}rieure des Postes et
T\'{e}l\'{e}graphes (today Sup'Telecom ParisTech) in 1908 proves that he had
applied his own concept of \textit{limit cycle} to a problem of Wireless Telegraphy preceding
thus Andronov of twenty years, it reopens the discussion of Poincar\'{e}'s
French legacy in Dynamical System Theory or, more precisely in Nonlinear
Oscillations Theory.

Poincar\'{e}'s ``forgotten lectures'' will be presented in the section B and
their reception in France before the First World War and in the 1920's in
the French engineers community in section C. Then, a special attention will
be paid to the role of Jean-Baptiste Pomey who asked \'{E}lie Cartan to
solve a problem of sustained oscillations in an electrotechnics oscillator
and, to the particular case of Alfred Li\'{e}nard who proved, under certain
conditions, existence and uniqueness of a periodic solution for such an
oscillator without having regarded this periodic solution as a
``Poincar\'{e}'s limit cycle''.

Starting from 1929, Andronov's note at the \textit{Comptes-Rendus} seems to have become the
reference in terms of connection with Poincar\'{e}'s works. So, the
reception of Andronov's results by the French scientific community will be
analyzed in section D in considering a selection of works published in
France between 1929 to 1943.

In the last section, the fact that Poincar\'{e}'s lectures have been
forgotten as well as the fact that neither Cartan nor Li\'{e}nard have made
any connection with Poincar\'{e}'s works although they have obviously used
some of them, will be discussed. Moreover, the role of Jacques Hadamard in
the diffusion in France of the works of the Russian schools and of
Poincar\'{e}'s methods will be also pointed out. Thus, the question of
Poincar\'{e}'s legacy will appear in a new perspective.

\end{abstract}

\section{Poincar\'{e}'s forgotten lectures on Wireless Telegraphy}

On July 4$^{th}$ 1902 Poincar\'{e} became Professor of Theoretical
Electricity at the \'{E}cole Sup\'{e}rieure des Postes et
T\'{e}l\'{e}graphes (today Sup'T\'{e}lecom) in Paris where he taught until
1910. The director of this school, \'{E}douard \'{E}stauni\'{e} (1862-1942),
also asked him to give a series of conferences every two years in May-June
from 1904 to 1912. He told about Poincar\'{e}'s first lecture of 1904:

\begin{quote}

\og{}D\`{e}s les premiers mots, il apparut que nous allions assister au
travail de recherche de cet extraordinaire et g\'{e}nial math\'{e}maticien
{\ldots} \`{A} chaque obstacle rencontr\'{e}, une courte pause marquait
l'embarras, puis d'un coup d'\'{e}paule, Poincar\'{e} semblait d\'{e}fier la
fonction g\^{e}nante {\ldots}\fg{}
\end{quote}

In 1908, Poincar\'{e} chose as the subject: wireless telegraphy. The text of
his lectures was first published weekly in the journal \textit{La Lumi\`{e}re \'{E}lectrique} \cite{Poin1908}
before being edited as a book the year after \cite{Poin1909}. In the
fifth and last part of these lectures entitled: T\'{e}l\'{e}graphie
dirig\'{e}e~: oscillations entretenues (Directive telegraphy: sustained
oscillations) Poincar\'{e} stated a necessary condition for the
establishment of a stable regime of sustained oscillations in the singing
arc (a forerunner device of the triode used in Wireless Telegraphy). More
precisely, he demonstrated the existence, in the phase plane, of a \textit{stable limit cycle}.

\subsection{The singing arc equation}

Starting from the following diagram (see Fig. 1), Poincar\'{e} \cite[p.
390]{Poin1908} explained that this circuit consists of an Electro Motive Force
(E.M.F.) of direct current E, a resistance R and a self-induction, and in
parallel, a singing arc and another self-induction L and a capacitor. In
order to provide the differential equation modeling the sustained
oscillations he calls $x$ the capacitor charge and $i$ the current in the external
circuit.

\begin{figure}[htbp]
\centerline{\includegraphics[width=12.36cm,height=4.63cm]{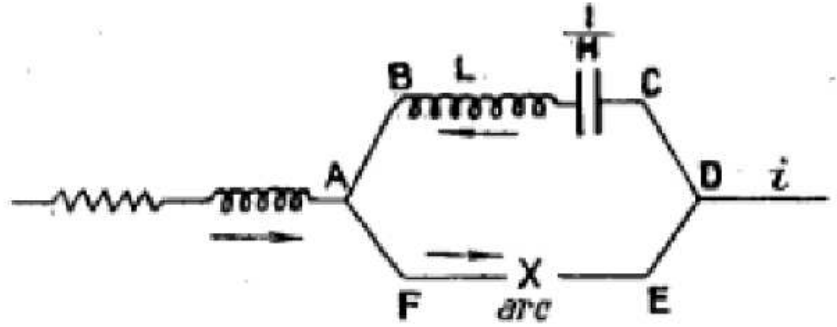}}
\label{fig1}
\caption{Circuit diagram of the singing arc, Poincar\'{e} \cite[p. 390]{Poin1908}}
\end{figure}

Thus, the current in the branch (ABCD) comprising the capacitor of capacity
$1 \mathord{\left/ {\vphantom {1 H}} \right. \kern-\nulldelimiterspace} H$
may be written: ${x}'={dx} \mathord{\left/ {\vphantom {{dx} {dt}}} \right.
\kern-\nulldelimiterspace} {dt}$. The current intensity $i_a $ in the branch
(AFED) comprising the singing may be written while using Kirchhoff's law:
$i_a =i+{x}'$. Then, Poincar\'{e} established the following second order
nonlinear differential equation for the sustained oscillations in the
singing arc:

\begin{equation}
\label{eq1}
L{x}''+\rho {x}'+\theta \left( {{x}'} \right)+Hx=0
\end{equation}

He specified that the term $\rho {x}'$ corresponds to the internal
resistance of the self and various damping while the term $\theta \left(
{{x}'} \right)$ represents the E.M.F. of the arc which is related to the
intensity by a function, unknown at that time.

\subsection{Stability condition for sustained oscillations and limit cycles}

Then, Poincar\'{e} established, twenty years before Andronov \cite{Andro1929}, that
the stability of the periodic solution of the above equation depends on the
existence of a closed curve, i.e. of a \textit{stable limit cycle} in the phase plane he has defined in
his memoirs \og{}Sur les Courbes d\'{e}finies par une \'{e}quation
diff\'{e}rentielle\fg{} \cite[p. 168]{Poin1886}. He posed:

\[
{x}'=\frac{dx}{dt}=y \mbox{ ; } dt=\frac{dx}{y} \mbox{ ; } {x}'' = \frac{dy}{dt}=\frac{ydy}{dx}
\]

Thus, equation (\ref{eq1}) becomes:

\begin{equation}
\label{eq2}
Ly\frac{dy}{dx}+\rho y+\theta \left( y \right)+Hx=0
\end{equation}

Poincar\'{e} \cite[p. 390]{Poin1908} stated then that:\\

\begin{quote}
``Sustained~oscillations~correspond\\ to~closed curves if there exist any.''\\
\end{quote}

and he gave the following representation for the solution of equation (\ref{eq2}):

\begin{figure}[htbp]
\centerline{\includegraphics[width=8cm,height=8cm]{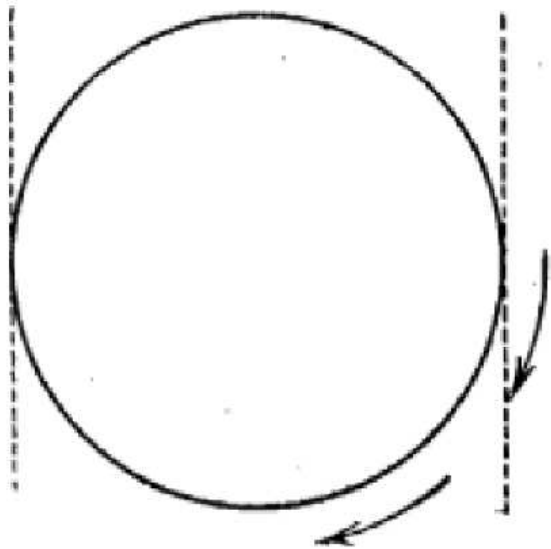}}
\label{fig2}
\caption{Closed curve solution of Eq. (\ref{eq2}), Poincar\'{e} [1908, p. 390]}
\end{figure}

Let's notice that this closed curve is only a \textit{metaphor} of the solution since
Poincar\'{e} does not use any graphical integration method such as
\textit{isoclines}.

Then, Poincar\'{e} explained that if $y=0$ then ${dy} \mathord{\left/
{\vphantom {{dy} {dx}}} \right. \kern-\nulldelimiterspace} {dx}$ is infinite
and so, the curve admits vertical tangents. Moreover, if $x$ decreases ${x}'$
i.e. $y$ is negative. He concluded that the trajectory curves turns in the
direction indicated by the arrow (see Fig. 2) and wrote:

\begin{quote}

``\textit{Stability condition. -- }Let's consider another non-closed curve satisfying the differential
equation, it will be a kind of spiral curve approaching indefinitely near
the closed curve. If the closed curve represents a stable regime, by
following the spiral in the direction of the arrow one should be brought
back to the closed curve, and provided that this condition is fulfilled the
closed curve will represent a stable regime of sustained waves and will give
rise to a solution of this problem.''
\end{quote}

Then, it clearly appears that the closed curve which represents a stable
regime of sustained oscillations is nothing else but a limit cycle as
Poincar\'{e} \cite[p. 261]{Poin1882} has introduced it in his own famous memoir ``On
the curves defined by differential equations'' and as Poincar\'{e} \cite[p. 25]{Poin1884} has later defined it in the notice on his own scientific works.

But this, first \textit{giant step} is not sufficient to prove the stability of the oscillating
regime. Poincar\'{e} had to demonstrate now that the periodic solution of
equation (\ref{eq1}) (the closed curve) corresponds to a \textit{stable limit cycle}.

\subsection{Possibility condition of the problem: limit cycle stability}

In the following part of his lectures, Poincar\'{e} gave what he calls a
\og{}condition de possibilit\'{e} du probl\`{e}me\fg{}. In fact, he established a
stability condition of the periodic solution of equation (\ref{eq1}), i.e. a
stability condition of the limit cycle under the form of inequality. After
multiplying equation (\ref{eq2}) by ${x}'dt$ Poincar\'{e} integrated it over one
period while taking into account that the first and fourth term vanish since
they correspond to the conservative part of this nonlinear equation. He
obtained:

\[
\rho \int {{x}'^2dt} +\int {\theta \left( {{x}'} \right){x}'dt=0}
\]

Then, he explained that since the first term is quadratic, the second one
must be negative in order to satisfy this equality. So, he stated that the
oscillating regime is stable iff:

\begin{equation}
\label{eq3}
\int {\theta \left( {{x}'} \right){x}'dt<0}
\end{equation}

As exemplified below, Poincar\'{e}'s approach is identical to which will be
used by Alfred Li\'{e}nard twenty years later.

\section{The reception of Poincar\'{e}'s lectures on Wireless Telegraphy in France}

The discovery of these Poincar\'{e}'s ``forgotten lectures'' on Wireless
Telegraphy implied the analysis of their influence on the French engineers
community.

\subsection{Before the First World War (1910-1914)}

During this period (1910-1914) one scientific reference to these conferences
could be found. It is in the book by Gaston \'{E}mile Petit and L\'{e}on Bouthillon
entitled\textit{ La T\'{e}l\'{e}graphie Sans Fil} published in 1910 in which one can read\footnote{See Petit {\&}
Bouthillon \cite[p. 128]{Petit1910}.}:

\begin{quote}
\og{}Le probl\`{e}me de la direction des ondes est donc r\'{e}soluble. Les
ondes peuvent th\'{e}oriquement \^{e}tre concentr\'{e}es en faisceau comme
les rayons lumineux par des dispositifs appropri\'{e}s ($^{1})$.\fg{}\\

($^{1})$ \footnotesize{Poincar\'{e}, Conf\'{e}rences sur la T\'{e}l\'{e}graphie sans fil
faites \`{a} l'\'{e}cole professionnelle sup\'{e}rieure des Postes et
T\'{e}l\'{e}graphes de Paris, 1908, p. 25.}
\end{quote}

Unfortunately, this unique quotation is very disappointing since it does not
refer to the part of Poincar\'{e}'s lectures concerning the sustained
oscillations. Nevertheless, since Petit (ESPT, 1906) and Bouthillon (ESPT,
1907) were students at the \'{E}cole Sup\'{e}rieure des Postes et
T\'{e}l\'{e}graphes (ESPT) when Poincar\'{e} was teaching there, one can
suppose that they could have attended his lectures of 1908.

Then, there are several allusions to his lectures in the eulogies made at
the time of Poincar\'{e}'s death on July 17$^{th}$ 1912 and after.

This is the case for example of the French engineer Andr\'{e} Blondel
(1863-1938) who wrote on July 27$^{th}$ a tribute to Poincar\'{e}:

\begin{quote}

\og{}De m\^{e}me aussi il avait \'{e}t\'{e} conduit \`{a} \'{e}tudier dans ses
\textit{Conf\'{e}rences \`{a} l'\'{E}cole Sup\'{e}rieure des Postes et T\'{e}l\'{e}graphes}
le probl\`{e}me de la propagation de l'\'{e}lectricit\'{e}, \`{a} propos
duquel il a d\'{e}velopp\'{e} les recherches de Kohlrausch et pouss\'{e}
plus loin ses r\'{e}sultats. De m\^{e}me il fut amen\'{e} \`{a}
s'int\'{e}ress\'{e} \`{a} la t\'{e}l\'{e}graphie sans fil, qui \'{e}tait
pour lui une application des th\'{e}ories qu'il avait d\'{e}velopp\'{e}es
sur les oscillations \'{e}lectriques\footnote{See Blondel \cite[p. 100]{Blondel1912}.}.\fg{}

\end{quote}

The year after, Gaston Darboux (1842-1917) in his historical praise
recalled\footnote{See Darboux \cite[p. 37]{Darboux1913}.}:

\begin{quote}

\og{}Les conf\'{e}rences qu'il a donn\'{e}es \`{a} l'\'{E}cole de
T\'{e}l\'{e}graphie nous montrent \'{e}galement combien il se tenait
pr\`{e}s de l'exp\'{e}rience, et quels services il a rendus aux praticiens.

L'\'{e}quation, dite des t\'{e}l\'{e}graphistes, nous fait conna\^{\i}tre,
comme on sait, les lois de la propagation d'une perturbation \'{e}lectrique
dans un fil. Poincar\'{e} int\`{e}gre cette \'{e}quation par une m\'{e}thode
g\'{e}n\'{e}rale qui peut s'appliquer \`{a} un grand nombre de questions
analogues. Le r\'{e}sultat varie suivant la nature du r\'{e}cepteur
plac\'{e} sur la ligne, ce qui se traduit math\'{e}matiquement par un
changement dans les \'{e}quations aux limites, mais la m\^{e}me m\'{e}thode
permet de traiter tous les cas.

Dans une seconde s\'{e}rie de conf\'{e}rences, Poincar\'{e} a \'{e}tudi\'{e}
le r\'{e}cepteur t\'{e}l\'{e}phonique ; un point qu'il a mis
particuli\`{e}rement en \'{e}vidence, c'est le r\^{o}le des courants de
Foucault dans la masse de l'aimant.

Enfin, dans une troisi\`{e}me s\'{e}rie de conf\'{e}rences, il a trait\'{e}
les diverses questions math\'{e}matiques relatives \`{a} la
t\'{e}l\'{e}graphie sans fil : \'{e}mission, champ en un point
\'{e}loign\'{e} ou rapproch\'{e}, diffraction, r\'{e}ception, r\'{e}sonance,
ondes dirig\'{e}es, \underline {ondes entretenues}~($^{1})$.\fg{}

($^{1})$ \footnotesize{Ces Conf\'{e}rences ont \'{e}t\'{e} publi\'{e}es dans la collection
des cours de 1'\'{e}cole et dans la revue \textit{L'\'{E}clairage \'{e}lectrique}.}

\end{quote}

\subsection{After the First World War (1919-1928)}

One might think that his lectures were forgotten because they dealt with an
application to a device for Wireless Telegraphy: the singing arc which had
become obsolete after the war.

\subsubsection{Janet and the electrical-mechanical analogy (1919)}

Of course, during the First World War the triode invented on January
15$^{th}$ 1907 by Lee de Forest (1873-1961) had supplanted the singing arc.
However, a French engineer named Paul Janet (1863-1937) published in 1919 a
note at the \textit{Comptes-Rendus} in which he established an analogy between the triode and the
singing arc. In this paper, entitled \og{}Sur une analogie \'{e}lectrotechnique des oscillations entretenues\fg{} and, by using the
classical electrical-mechanical analogy, Janet \cite{Janet1919} provided the nonlinear
differential equation characterizing the oscillations sustained by the
singing arc and by the triode:

\begin{equation}
\label{eq4}
L\frac{d^2i}{dt^2}+\left[ {R-{f}'\left( i \right)}
\right]\frac{di}{dt}+\frac{k^2}{K}i=0
\end{equation}

By using the electrical-mechanical analogy and by posing $H\to {k^2}
\mathord{\left/ {\vphantom {{k^2} K}} \right. \kern-\nulldelimiterspace} K$
and $\theta \left( {{x}'} \right)={f}'\left( i \right)\frac{di}{dt}$
it is easy to show that Eq. (\ref{eq1}) and Eq. (\ref{eq4}) are completely analogous.
Nevertheless, there is no reference to Poincar\'{e} in this note.

\subsubsection{Pomey and the singing arc equation (1920)}

On June 28$^{th}$ 1920, a textbook written by the engineer Jean-Baptiste
Pomey (1861-1943) entitled: ``~Introduction \`{a} la th\'{e}orie des
courants t\'{e}l\'{e}phoniques et de la radiot\'{e}l\'{e}graphie~'' was
published in France. A former student of the \'{E}cole Sup\'{e}rieure des
Postes et T\'{e}l\'{e}graphes, Pomey (ESPT, 1883) became Professor of
Theoretical Electricity in this school alongside Henri Poincare and then
director from 1924 to 1926. In Chapter XIX of his book, devoted to the
generation of sustained oscillations Pomey \cite[p. 375]{Pomey1920} wrote:

\begin{quote}

\og{}Pour que des oscillations soient engendr\'{e}es spontan\'{e}ment et
s'entretiennent, il ne suffit pas que l'on ait un mouvement p\'{e}riodique,
il faut encore que ce mouvement soit stable.\fg{}

\end{quote}

Then, he provided the nonlinear differential equation of the singing arc:

\begin{equation}
\label{eq5}
L{x}''+R{x}'+\frac{1}{C}x=E_0 +a{x}'-b{x}'^3
\end{equation}

By posing $H=1 \mathord{\left/ {\vphantom {1 C}} \right.
\kern-\nulldelimiterspace} C$, $\rho =R$ and $\theta \left( {{x}'}
\right)=-E_0 -a{x}'+b{x}'^3$ it is obvious that Eq. (\ref{eq1}) and Eq. (\ref{eq5}) are
completely identical\footnote{For more details see Ginoux {\&} Lozi
\cite{GinLoz2012}.}. Moreover, it is striking to observe that Pomey has used exactly
the same variable ${x}'$ as Poincar\'{e} to represent the current intensity.
Here again, there is no reference to Poincar\'{e}. This is very surprising
since Pomey was present during the last lecture of Poincar\'{e} at the
\'{E}cole Sup\'{e}rieure des Postes et T\'{e}l\'{e}graphes in 1912 whose he
had written the introduction. So, one can imagine that he could have
attended the lecture of 1908.

\subsubsection{\'{E}lie and Henri Cartan and the existence of a periodic solution (1925)}

On September 28$^{th}$ 1925, Pomey wrote a letter to the mathematician
\'{E}lie Cartan (1869-1951) in which he asked him to provide a condition for
which the oscillations of an electrotechnics device analogous to the singing
arc and to the triode whose equation is exactly that of Janet (see Eq. (\ref{eq4}))
are sustained. Within ten days, \'{E}lie Cartan and his son Henri sent an
article entitled: \og{}Note sur la g\'{e}n\'{e}ration des oscillations
entretenues\fg{} in which they proved the existence of a periodic solution for
Janet's equation (\ref{eq4}). In fact, their proof is based on a diagram (see Fig.
3) which corresponds exactly to a ``first return map'' diagram introduced by
Poincar\'{e} in his memoir \og{}Sur les Courbes d\'{e}finies par une
\'{e}quation diff\'{e}rentielle\fg{} \cite[p. 251]{Poin1882}. They wrote:

\begin{quote}

\og{}Les points H$_{1}$, H$_{2}$, H$_{3}$ o\`{u} la courbe coupe la
bissectrice correspondent \`{a} des solutions \textit{p\'{e}riodiques} (oscillations entretenues)
\textit{dont l'existence est ainsi d\'{e}montr\'{e}e}. Elles sont p\'{e}riodiques parce qu'en partant d'un minimum
donn\'{e}$\mbox{ }-i_1 $ , le maximum suivant est \'{e}gal \`{a} $i_1 $, par
suite le minimum suivant e \`{a} $-i_1 $, etc. On peut maintenant voir
facilement que \textit{toute solution tend vers une solution p\'{e}riodique}\footnote{See Cartan \cite[p. 1199]{Cartan1925}.}.\fg{}

\end{quote}

\begin{figure}[htbp]
\centerline{\includegraphics[width=10cm,height=8.6cm]{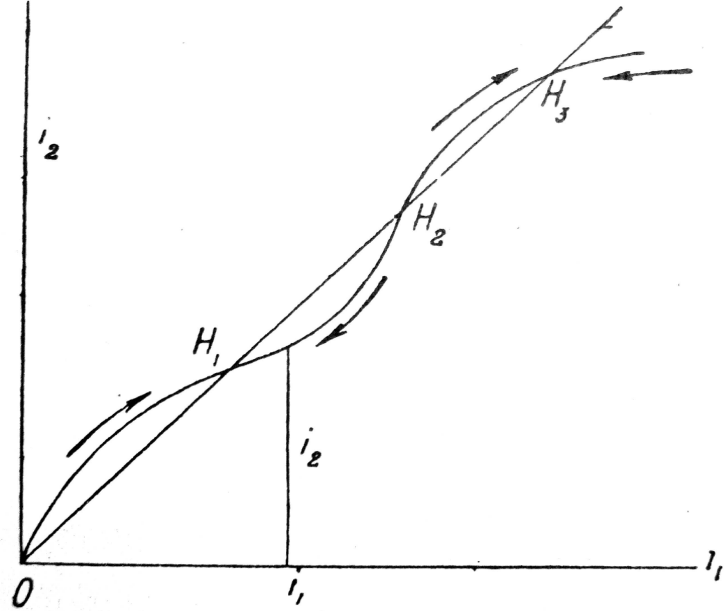}}
\label{fig3}
\caption{``First return map'' diagram, Cartan [1925, p. 1199]}
\end{figure}

Obviously, Fig. 3 exhibits an application of Poincar\'{e}'s method although
there is no reference to his works. Moreover, Cartan didn't recognize in
this periodic solution a limit cycle of Poincar\'{e}.

\subsubsection{Li\'{e}nard and the existence and uniqueness of a periodic solution (1928)}

Three years after, on May 1928, the engineer Alfred Li\'{e}nard (1869-1958)
published an article entitled \og{}\'{E}tude des oscillations entretenues\fg{}
in which he studied the solution of the following nonlinear differential
equation:

\begin{equation}
\label{eq6}
\frac{d^2x}{dt^2}+\omega f\left( x \right)\frac{dx}{dt}+\omega ^2x=0
\end{equation}

Such an equation is a generalization of the well-known Van der Pol's
equation and of course of Janet's equation (\ref{eq4}). Under certain assumptions on
the function $F\left( x \right)=\int_0^x {f\left( x \right)dx} $ less
restrictive than those chosen by Cartan \cite{Cartan1925} and Van der Pol \cite{VdP1926},
Li\'{e}nard \cite{Lien1928} proved the existence and uniqueness of a periodic
solution of Eq. (\ref{eq6}). Then, Li\'{e}nard \cite[p. 906]{Lien1928} plotted this solution
(see Fig. 4) and wrote:

\begin{quote}

\og{}On se rend ainsi compte que la courbe int\'{e}grale d\'{e}crit une sorte
de spirale tendant asymptotiquement vers la courbe ferm\'{e}e D. Pour les
courbes int\'{e}grales ext\'{e}rieures \`{a} la courbe ferm\'{e}e, c'est
OA$_{2}$ qui devient inf\'{e}rieur \`{a} OA$_{1}$. La courbe se rapproche
encore de la courbe D, mais par l'ext\'{e}rieur. En raison de ce~fait que
toutes les courbes int\'{e}grales, int\'{e}rieures ou ext\'{e}rieures,
parcourues dans le sens des temps croissants, tendent asymptotiquement vers
la courbe D on dit que le mouvement p\'{e}riodique correspondant est un
mouvement \textit{stable}.\fg{}

\end{quote}

\begin{figure}[htbp]
\centerline{\includegraphics[width=6cm,height=9.3cm]{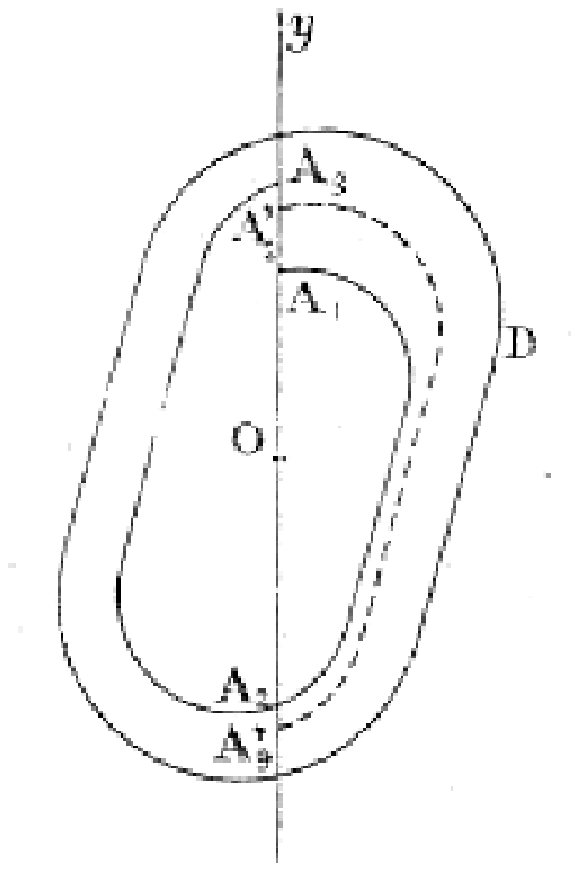}}
\label{fig4}
\caption{Closed curve solution of Eq. (\ref{eq6}), Li\'{e}nard [1928, p. 905].}
\end{figure}

The Li\'{e}nard \cite[p. 906]{Lien1928} explained that the condition for which the
``periodic motion'' is stable is given by the following inequality:

\begin{equation}
\label{eq7}
\int\limits_\Gamma {F\left( x \right)dy} >0
\end{equation}

By considering that the trajectory curve describes the closed curve
clockwise in the case of Poincar\'{e} and counter clockwise in the case of
Li\'{e}nard, it is easy to show that both conditions (\ref{eq3}) and (\ref{eq7}) are
completely identical\footnote{For more details see Ginoux \cite{Gin2011} and Ginoux
\cite{Gin2012}.} and represents an analogue of what is now called ``orbital
stability''. Again, one can find no reference to Poincar\'{e} in
Li\'{e}nard's paper. Moreover, it is very surprising to observe that he
didn't used the terminology ``limit cycle'' to describe its periodic
solution. The case of Li\'{e}nard is really a riddle that we will discussed
below.

\section{The reception of Andronov's results in France}

After the Poincar\'{e}'s death in 1912, the French mathematician Jacques
Hadamard (1865-1963) succeeded him at the Academy of Sciences. According to
Maz'ya et Shaposhnikova \cite[p.~181]{Mazya2005} Hadamard was elected corresponding
member of the Russian Academy of Sciences in 1922 and, foreign member of the
Academy of Sciences of the USSR in 1929. That's probably the reason why he
was asked to review\footnote{The original of this note is presented in
Ginoux \cite[p. 494-496]{Gin2011}.} and to present at the \textit{Comptes-Rendus} on October 14$^{th}$
1929 a note from the young Russian mathematician Aleksandr' Aleksandrovich
Andronov (1901-1952) entitled:

\begin{quote}
\og{}Les cycles limites de Poincar\'{e} et la th\'{e}orie des oscillations
auto-entretenues.\footnote{Andronov \cite{Andro1929}. It has been pointed out by
Ginoux \cite[p. 176]{Gin2011} that this note has been preceded by a presentation at
the Congress of Russian Physicists between 5$^{th}$ and 16$^{th}$ August
1928. See Andronov \cite{Andro1928}.}\fg{}
\end{quote}

In this short paper (only three pages), Andronov \cite{Andro1929} stated a
correspondence between the periodic solution of self-oscillating systems
representing for example the oscillations of a radiophysics device used in
Wireless Telegraphy and the concept of stable \textit{limit cycle} introduced by Poincar\'{e} \cite[p. 261]{Poin1882}.

In order to analyze the reception of Andronov's result, a selection of works
published in France and elsewhere during the period (1930-1943) will be
briefly presented.

\subsection{Van der Pol's relaxation oscillations and limit cycles (1930)}

Looking at the famous plot of the graphical integration of the triode
oscillator exhibited by Van der Pol \cite[p. 983]{VdP1926} (see Fig. 5), one might
think that he has immediately recognized that such a periodic solution was a
\textit{stable limit cycle} of Poincar\'{e}.

\begin{figure}[htbp]
\centerline{\includegraphics[width=8cm,height=10cm]{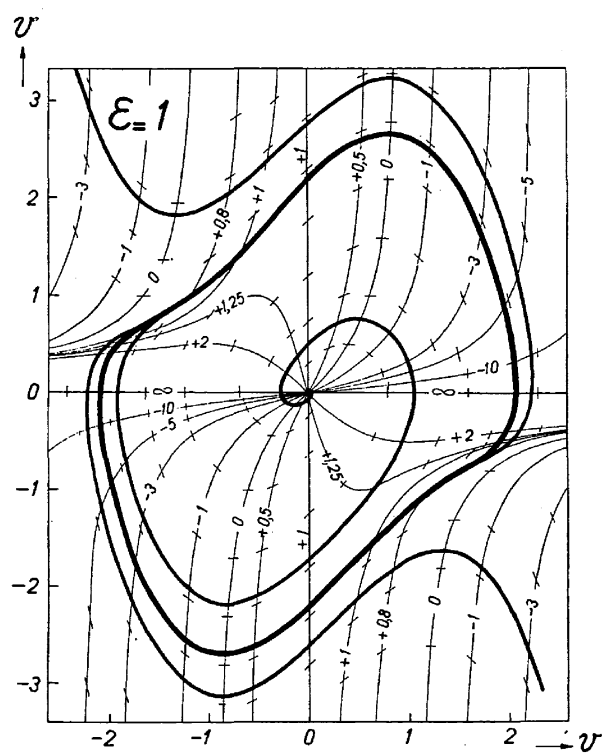}}
\caption{Graphical integration of the triode oscillator, Van der Pol \cite[p. 983]{VdP1926}}
\label{fig5}
\end{figure}

This is not the case. It is just after the publication of Andronov's note at
the \textit{Comptes-Rendus} that Van der Pol realized that the periodic solution he has plotted
(see Fig. 5) was a limit cycle. It was during a lecture given in Paris at
the \'{E}cole Sup\'{e}rieure d'\'{E}lectricit\'{e} on March 10$^{th}$ and
11$^{th}$ 1930. Van der Pol \cite[p. 294]{VdP1930} reproduced the figures of his
original paper of 1926 (comprising the Fig. 5) and he said:

\begin{quote}

\og{}On remarque sur chacune de ces trois figures une courbe int\'{e}grale
ferm\'{e}e~; c'est un exemple de ce que Henri POINCAR\'{E} a appel\'{e} un
\textit{cycle limite} ($^{1})$, parce que les courbes int\'{e}grales s'en rapprochent
asymptotiquement.\fg{}

($^{1})$ \footnotesize{Voir par exemple~: \textit{A. Andronow}, Les cycles limites de Poincar\'{e} et la
th\'{e}orie des oscillations auto-entretenues, C. R. \textbf{189}, 559
(1929).}

\end{quote}

Let's notice that although he quotes Poincar\'{e}, he makes reference to
Andronov. Moreover, he also quotes in the following, the papers of Cartan \cite{Cartan1925} and
that of Li\'{e}nard \cite{Lien1928} but it does not seem that he has ever
used their works. During his visits in Paris, Van der Pol was hosted by a
young French engineer named Philippe Le Corbeiller (1891-1980) who helped
him to translate his talks\footnote{See Van der Pol \cite[p. 312]{VdP1930}.}. Le
Corbeiller who was probably present at the \'{E}cole Sup\'{e}rieure
d'\'{E}lectricit\'{e} on March 10$^{th}$ and 11$^{th}$ 1930 was invited to
give a lecture at the \textit{Third International Congress of Applied Mechanics} held in Stockholm from 24$^{th}$ to 29$^{th}$ August
1930. He was accompanied by Van der Pol himself and by Alfred Li\'{e}nard.

\subsection{Le Corbeiller and the Theory of Nonlinear Oscillations (1930)}

During his talk entitled \og{}Sur les oscillations des r\'{e}gulateurs\fg{} Le
Corbeiller \cite[p. 211]{LeCorb1931a} recalled:

\begin{quote}

\og{}Si nous savons que le syst\`{e}me machine-r\'{e}gulateur pr\'{e}sente
effectivement des oscillations p\'{e}riodiques, cela signifiera que parmi
les courbes int\'{e}grales trac\'{e}es sur la surface caract\'{e}ristique il
y en a au moins \textit{une} qui est une courbe ferm\'{e}e. Mais le syst\`{e}me
n'\'{e}tant plus lin\'{e}aire \`{a} coefficients constants, les solutions
infiniment voisines ne seront plus homoth\'{e}tiques \`{a} cette courbe,
mais s'en approcheront asymptotiquement, c'est-\`{a}-dire que la solution
p\'{e}riodique correspondra \`{a} un \textit{cycle limite} de POINCAR\'{E}, comme l'a fait
remarquer M. ANDRONOW. Son amplitude sera ainsi bien d\'{e}termin\'{e}e.\fg{}

\end{quote}

Then, he ended his article by this sentence:

\begin{quote}

\og{}Je ne puis que renvoyer aux remarquables travaux de cet auteur\footnote{
Van der Pol.}, auxquels M. LI\'{E}NARD et M. ANDRONOW ont apport\'{e} des
compl\'{e}ments fort int\'{e}ressants.\fg{}

\end{quote}

As an Historian of Sciences, Le Corbeiller presented during his various
lectures a synthesis of the different results obtained in the field of
relaxation oscillations. Moreover, his contribution for the understanding of
the processes of development of the nonlinear oscillations theory is
fundamental since he recalls the essential steps that would surely have
fallen into oblivion without his intervention. Nevertheless, it is striking
to notice that in his famous lecture given in Paris at the Conservatoire
National des Arts {\&} M\'{e}tiers on May 6$^{th}$ and on May 7$^{th}$ 1931,
Le Corbeiller \cite[p. 4]{LeCorb1931b} quoted very few Andronov and never Poincar\'{e}:

\begin{quote}

\og{}Mais c'est un physicien hollandais, M. Balth. van der Pol, qui, par sa
th\'{e}orie des \textit{oscillations de relaxation} (1926) a fait avancer la question d'une mani\`{e}re
d\'{e}cisive. Des savants de divers pays travaillent actuellement \`{a}
\'{e}largir la voie qu'il a trac\'{e}e~; de ces contributions, la plus
importante nous para\^{\i}t \^{e}tre celle de M. Li\'{e}nard (1928). Des
recherches math\'{e}matiques fort int\'{e}ressantes sont poursuivies par M.
Andronow, de Moscou.\fg{}

\end{quote}

On April 22$^{nd}$ 1932, Le Corbeiller gave a lecture in Paris at the
\'{E}cole Sup\'{e}rieure des Postes et T\'{e}l\'{e}graphes where he has been
student many years before (ESPT, 1914). Let's recall that Poincar\'{e}
taught in this school from 1902 till 1912 where he also gave many lectures
and in particular his ``forgotten lectures'' on Wireless Telegraphy.

Discussing the graphical integration of the triode oscillator proposed by
Van der Pol \cite[p. 983]{VdP1926} (See Fig. 5. above), Le Corbeiller \cite[p.
708-709]{LeCorb1932} wrote:

\begin{quote}

\og{}La th\'{e}orie g\'{e}n\'{e}rale de ces courbes int\'{e}grales ferm\'{e}e,
ou \textit{cycles limites}, a \'{e}t\'{e} faite par H. Poincar\'{e} ($^{2})$ [1]~; la
d\'{e}monstration de l'existence d'un cycle limite unique, dans ce cas
actuel, est due \`{a} M. Li\'{e}nard [16].

($^{2})$ D'une mani\`{e}re tout \`{a} fait g\'{e}n\'{e}rale, l'\'{e}quation
$\dfrac{dx}{X\left( {x,y} \right)}=\dfrac{dy}{Y\left( {x,y} \right)}$
\'{e}quivaut aux deux suivantes~:

\[
\dfrac{d^2x}{dt^2}-y\left( {\dfrac{Y}{X}+x} \right)+x=0\mbox{, } y=\dfrac{dx}{dt}.
\]

Le m\'{e}moire cit\'{e} de Poincar\'{e} \'{e}quivaut donc \`{a} l'\'{e}tude du syst\`{e}me oscillant conservatif
$\dfrac{d^2x}{dt^2}+x=0$, soumis \`{a} des forces de dissipation et d'entretien dont la r\'{e}sistance est une force \textit{quelconque} de $x$ et de
$\dfrac{dx}{dt}$~:\fg{}

\[
F\left( {x,\frac{dx}{dt}} \right)=y\frac{Y}{X}+x.
\]

\footnotesize{[1] H. POINCAR\'{E}, M\'{e}moire sur les courbes d\'{e}finies par une
\'{e}quation diff\'{e}rentielle, Deuxi\`{e}me partie, \textit{Journal de Math\'{e}m. Pures et app.} 8, 251, 1882~; et
\textit{{\OE}uvres}, T. 1, p. 44.}\\
\footnotesize{[16] A. LI\'{E}NARD, \'{E}tude des oscillations entretenues, \textit{Rev. g\'{e}n. d'\'{e}lectr.}, 901 et 946, 1926.}

\end{quote}

It is very interesting to remark that Le Corbeiller made reference to the
original paper of Poincar\'{e} and that he also gave many mathematical
details on the way of writing the differential equation characterizing the
oscillations of a dissipative system.

\subsection{The Li\'{e}nard riddle (1931)}

As recalled above, in his first paper entitled \og{}\'{E}tude des oscillations
entretenues\fg{}, Li\'{e}nard \cite{Lien1928} proved the existence and uniqueness of a
periodic solution of a generalized Van der Pol's equation without making any
connection with Poincar\'{e}'s works. Then, less than one year after the
presentation of Andronov's notes at the \textit{Comptes-Rendus}, Li\'{e}nard participated with Le
Corbeiller and Van der Pol to the \textit{Third International Congress of Applied Mechanics} held in Stockholm from 24$^{th}$ to
29$^{th}$ August 1930 where he presented an article entitled:
\og{}Oscillations auto-entretenues\fg{}. According to the title, one might have
thinking that, in this second and last publication on this subject,
Li\'{e}nard would have taken account of Andronov's result and that he would
have established a connection between the periodic solution and a
Poincar\'{e}'s limit cycle. But, surprisingly not.

In this work, Li\'{e}nard \cite{Lien1931} first summarized his previous results and
then he generalized another result established by Andronov and Witt \cite{Andro1930}
in their second and last note at the \textit{Comptes-Rendus} in which they studied the ``Lyapunov
stability'' of the periodic solution, $i.e.$ the stability of a \textit{limit cycle} or ``orbital
stability''. In order to extend Andronov and Witt's proposition, Li\'{e}nard \cite[p. 176]{Lien1931} made use of ``variational equations'', $i.e.$ of a method introduced by Poincar\'{e} \cite[p. 162]{Poin1882} in the first volume of his famous
\og{}M\'{e}thodes Nouvelles de la M\'{e}canique C\'{e}leste\fg{} and which
corresponds to what is today known under the name of the computation of
``characteristics exponents''. To do that, he modified his own equation (\ref{eq6})
and replaced it by the following which is now know as ``Li\'{e}nard's
equation'':

\begin{equation}
\label{eq8}
\frac{d^2x}{dt^2}+\omega f\left( {x,\frac{dx}{dt}} \right)+\omega ^2x=0
\end{equation}

Then, he wrote:

\begin{quote}

\og{}Si l'\'{e}quation (\ref{eq8}) admet une solution p\'{e}riodique, de p\'{e}riode
T, la condition pour que cette solution soit stable est que l'int\'{e}grale
pendant une p\'{e}riode de $\dfrac{\partial f\left( {x,{x}'}
\right)}{\partial {x}'}dt$ soit positive. La proposition, \'{e}tablie par
Messieurs ANDRONOW et WITT (\footnote{Li\'{e}nard quotes Andronov and Witt
[1930].}) dans le cas particulier o\`{u} la fonction $f\left( {x,{x}'}
\right)$ est tr\`{e}s petite se g\'{e}n\'{e}ralise imm\'{e}diatement.\fg{}

\end{quote}

Thus, Li\'{e}nard \cite{Lien1931} generalizes the result of Andronov and Witt \cite{Andro1930}
for the stability of a periodic solution, $i.e.$ of a limit cycle according to
Poincar\'{e}'s method of ``characteristics exponents'' but without quoting
Poincar\'{e}'s works and without using the terminology ``limit cycle'' for
describing the stable periodic solution\footnote{For more details see
Ginoux \cite[p. 210]{Gin2011}.}. However, this expression appears in the very first
pages of the article of Andronov and Witt \cite[p. 256]{Andro1930} in footnote:

\begin{quote}

\og{}($^{2})$ Pour la d\'{e}finition des auto-oscillations et la discussion du
cas d'un degr\'{e} de libert\'{e} voir A. ANDRONOW, \textit{Les cycle limites de Poincar\'{e} et la th\'{e}orie des oscillations auto-entretenues} (Comptes rendus, 189,
1929, p. 559).\fg{}

\end{quote}

Therefore, it seems very difficult to explain the attitude of Li\'{e}nard
especially since at the first International Conference on Nonlinear
Oscillations, to which he was invited, the question of periodic solutions of
type limit cycle has been much discussed.

\subsection{The first ``lost'' International Conference on Nonlinear Oscillations (1933)}

From 28 to 30 January 1933 the first International Conference of Nonlinear
Oscillations was held at the Institut Henri Poincar\'{e} (Paris) organized
at the initiative of the Dutch physicist Balthasar Van der Pol and of the
Russian mathematician Nikola\"{\i} Dmitrievich Papaleksi. The discovery of
this event, of which virtually no trace remains, was made possible thanks to
the report written by Papaleksi at his return in USSR. This document has
revealed, on the one hand, the list of participants who included French
mathematicians: Alfred Li\'{e}nard, \'{E}lie and Henri Cartan, Henri
Abraham, Eug\`{e}ne Bloch, L\'{e}on Brillouin, Yves Rocard, ... and, on the
other hand the content of presentations and discussions. The analysis of the
minutes of this conference highlights the role and involvement of the French
scientific community in the development of the theory of nonlinear
oscillations\footnote{For more details see Ginoux \cite{Gin2011} and Ginoux \cite{Gin2012}.}.

According to Papaleksi \cite[p. 211]{Papaleksi1934}, during his talk, Li\'{e}nard recalled
the main results of his study on sustained oscillations:

\begin{quote}

``Starting from its graphical method for constructing integral curves of
differential equations, he deduced the conditions that must satisfy the
nonlinear characteristic of the system in order to have periodic
oscillations, that is to say for that the integral curve to be a closed
curve, i.e. a limit cycle.''

\end{quote}

This statement on Li\'{e}nard must be considered with great caution. Indeed,
one must keep in mind that Papaleksi had an excellent understanding of the
work of Andronov \cite{Andro1929} and that his report was also intended for members of
the Academy of the USSR to which he must justified his presence in France at
this conference in order to show the important diffusion of the Soviet work
in Europe. Despite the presence of MM. Cartan, Lienard, Le Corbeiller and
Rocard it does not appear that this conference has generated, for these
scientists, a renewed interest in the problem of sustained oscillations and
limit cycles. However, although the theory of nonlinear oscillations does
not seem to be in France at that time a research priority, it is the subject
of several Ph-D theses and monographs discussed below.

\subsection{The French Ph-D theses}

During the period (1936-1943) several Ph-D theses were defended in France on
a subject strongly related to nonlinear oscillations. Two of them are
briefly recalled below\footnote{For instance four Ph-D thesis have been
found during this period and completely analyzed by Ginoux \cite{Gin2011}.}.

The first is that of R. Morched-Zadeh\footnote{No biographic information
could be found concerning this student except the fact that he was Iranian
but not parent with Lotfi Morched-Zadeh (personal communication). } who
defended a Ph-D thesis at the Facult\'{e} des Sciences de l'Universit\'{e}
de Toulouse in October 1936 entitled: \og{}\'{E}tude des oscillations de
relaxation et des diff\'{e}rents modes d'oscillations d'un circuit
comprenant une lampe n\'{e}on\fg{}. In the introduction of his study
Morched-Zadeh \cite[p. 3]{Zadeh1936} wrote:

\begin{quote}

\og{}Au point de vue th\'{e}orique, les cycles limites de H. POINCAR\'{E}
prennent une grande place dans la th\'{e}orie des oscillations
autoentretenues comme l'ont d\'{e}montr\'{e} A. ANDRONOW et A. WITT.\fg{}

\end{quote}

In this case it is very surprising to observe that Morched-Zadeh made
reference to the second and last note of Andronov and Witt \cite{Andro1930} at the
\textit{Comptes-Rendus} and not to the first which seemed to be better known and most quoted.

The aim of his work is an experimental study of relaxation oscillations of a
neon lamp submitted to various oscillating regimes comprising of course the
case of self-sustained oscillations. This led him to take the very first
pictures of a \textit{limit cycle} on a cathode ray tube oscilloscope (see Fig. 6.)

\begin{figure}[htbp]
\centerline{\includegraphics[width=7.75cm,height=10cm]{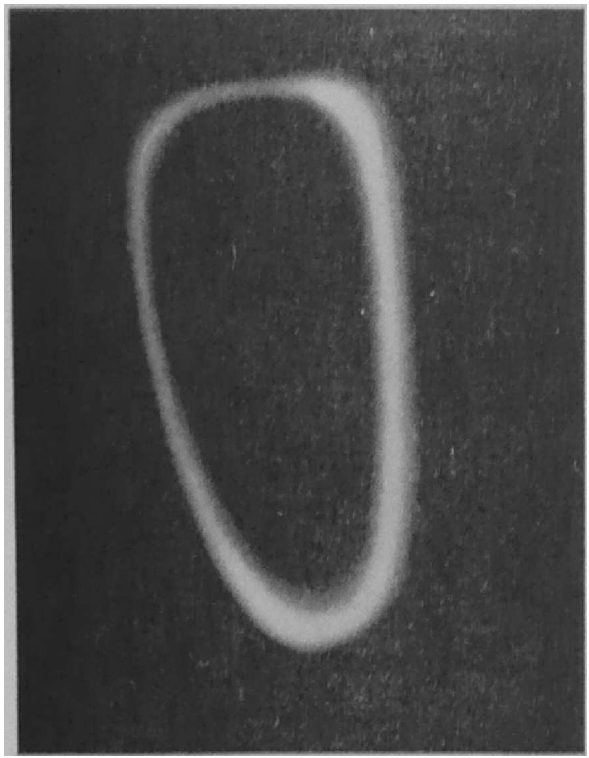}}
\label{fig6}
\caption{Limit cycle of the neon lamp oscillator, Morched-Zadeh [1936, p. 127]}
\end{figure}

The second is that of Jean Abel\'{e} (1886-1961) who was physicist,
philosopher and writer\footnote{ For biographic details see for example Ginoux \cite[p. 325]{Gin2011}.}.
He defended his Ph-D thesis entitled: ``~\'{E}tude
d'un syst\`{e}me oscillant entretenu \`{a} amplitude autostabilis\'{e}e et
application \`{a} l'entretien d'un pendule \'{e}lastique~'' at the
Facult\'{e} des Sciences de l'Universit\'{e} de Paris in front of a jury
comprising Yves Rocard. In the introduction of his work Abel\'{e} \cite[p.
18]{Abelé1943} wrote:

\begin{quote}

\og{}\`{A} un mouvement p\'{e}riodique \textit{stable} correspond une courbe int\'{e}grale
ferm\'{e}e \textit{dont s'approchent asymptotiquement en spirales}, \textit{de l'int\'{e}rieur et de l'ext\'{e}rieur}, \textit{pour t croissant}, \textit{les solutions voisines}. Un des probl\`{e}mes fondamentaux de la th\'{e}orie non
lin\'{e}aire consiste dans la recherche de ces courbes ferm\'{e}es, dites
\textit{cycles limites}\footnote{Abel\'{e} quotes Andronov \cite{Andro1929}.}.\fg{} \cite[p.
18]{Abelé1943}

\end{quote}

Although, the definition of a stable limit cycle exactly corresponds to that
given by Poincar\'{e} himself, Abel\'{e} quotes Andronov.

\subsection{Rocard's textbooks}

During the Second World War, the physicist Yves Rocard (1903-1992) published
two manuscripts. If the title of the first one \og{}Th\'{e}orie des Oscillateurs\fg{}
is very close to that of Andronov and Khaikin \cite{Andro1937}
published in Russian and entitled: ``{\cyr Teoriya kolebani{\u i}}\footnote{``Theory
of oscillations''.}'' the content is quite different. Rocard \cite{Rocard1941}
proposes a synthesis of several works done in this area as well as a summary
of the article of Van der Pol \cite{VdP1926} on relaxation oscillations with figures
including the Fig. 5 that he commented thus:

\begin{quote}

\og{}On voit au fur et \`{a} mesure que $\varepsilon $ cro\^{\i}t, se
d\'{e}former le cycle limite et appara\^{\i}tre les harmoniques.\fg{}

\end{quote}

This is the single occurrence of the terminology limit cycle in the whole
textbook, which is given without any reference\footnote{The question of the
absence of references in Rocard's book has been much discussed. Interviewed
on this issue, Rocard said that because of the war he was unable to access
these documents. In fact it has been shown in Ginoux \cite[p. 263]{Gin2011} that it
was not true.}.

In 1943, Rocard \cite{Rocard1943} published his \og{}Dynamique G\'{e}n\'{e}rale des
Vibrations\fg{} which has been wrongly considered as a textbook on nonlinear
oscillations. In fact, in this book which comprises sixteen chapters only
three deals with this subject. In chapter XV, Rocard \cite[p. 220]{Rocard1943} recalls
the results of Li\'{e}nard \cite{Lien1928} and plots the following figure:

\begin{figure}[htbp]
\centerline{\includegraphics[width=8cm,height=9.27cm]{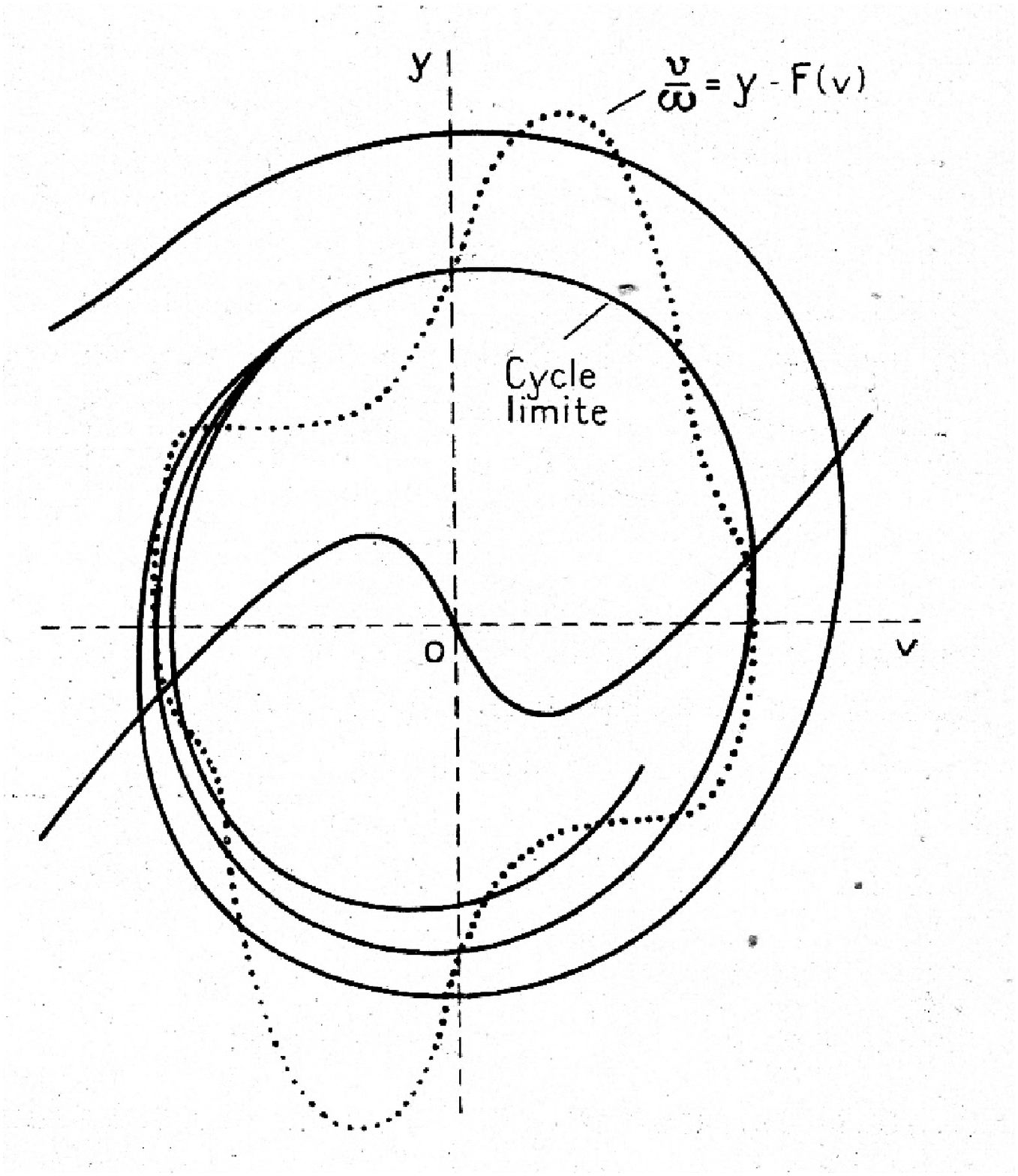}}
\label{fig7}
\caption{Limit cycle of relaxation oscillator, Rocard \cite[p. 220]{Rocard1943}}
\end{figure}

Then, he explained:

\begin{quote}

\og{}{\ldots}on constate (comme autrefois H. Poincar\'{e} l'a
d\'{e}montr\'{e}) que la courbe int\'{e}grale s'enroule un certain nombre de
fois et tend vers une courbe ferm\'{e}e dite \textit{cycle limite}, qui dans cette
repr\'{e}sentation correspond au r\'{e}gime d'oscillations permanent.\fg{}

\end{quote}

Here again, there is no reference neither to Poincar\'{e} nor to Andronov.

\section{Discussion}

This study has shown that two major contributions in the development of
nonlinear oscillation theory had occurred in France during the first half of
the XX$^{th}$ century. The first is the correspondence between the concept
of limit cycle and the existence of a stable regime of sustained
oscillations in Wireless Telegraphy established by Henri Poincare in 1908 in
these ``forgotten lectures'' at the \'{E}cole Sup\'{e}rieure des Postes et
T\'{e}l\'{e}graphes, and the second is the same kind of correspondence,
established twenty years later in a more general context by the Russian
mathematician Aleksandr' Andronov in his famous note at the
\textit{Comptes}-\textit{Rendus}.

If, the ``discovery'' of these ``forgotten lectures'' demonstrated that
Poincar\'{e} has stated that the periodic solution of a nonlinear
differential equation characterizing the nonlinear oscillations of a
particular radiophysics device named \textit{singing arc} is nothing else but a stable limit
cycle, it has really produced no reaction on the French scientific
community.

Nevertheless, the analysis of the influence of Andronov's note on this
scientific community from 1929 to 1943 has shown that it was nearly the
same. Li\'{e}nard, for unknown reasons didn't make use of the terminology
``limit cycle'' neither before 1929 nor after. Moreover, although he became
probably aware of Andronov's correspondence during the first ``lost''
International Conference on Nonlinear Oscillations in 1933 he has not
pursued his research in this area. But, in 1933, Li\'{e}nard was 64 years
old and near to retirement. This was not the case for Le Corbeiller who has
been one of the first to establish a deep connection with Poincar\'{e}'s
works. However, when the Second World War was declared he went to the USA
and became a Professor in Harvard. Concerning Rocard, who made only few
allusions to Poincar\'{e}'s concept of limit cycle in his textbooks, he
turned to nuclear research immediately after WWII.

Thus, it seems that although France has been a kind of crossroads for the
development of nonlinear oscillations theory, nobody has succeeded in
unifying French scientists around a research program in this area.

In fact, the mathematician Jacques Hadamard has been deeply involved in this
task at various levels. First, he has presented during the 1930's many notes
at the French Academy of Sciences on this subject coming from USSR: that of
Andronov, of Andronov and Witt but also that of Kryloff and Bogoliouboff.
But, he has also presented the works of Poincar\'{e} during his seminar at
the \textit{Coll\`{e}ge de France} from 1919 to 1932. Unfortunately it didn't provide any reaction from
the French scientific community.

So, although this community has produced many fundamental results necessary
for the development of the nonlinear oscillation theory such as that of
Li\'{e}nard for example there has been no research program like in USSR or
in USA and so, no ``School of nonlinear'' in France. The terrible impact of
the WWI and WWII is probably responsible for such a lack of organization.

During the commemoration of the centenary of Poincar\'{e}'s birth, the
newspaper \textit{Le Monde} published on May 15$^{th}$, 1954 an article entitled \og{}Les
conferences de Henri Poincar\'{e} \`{a} l'\'{E}cole Sup\'{e}rieure des
P.T.T.\fg{} (See Fig. 8) in which we learn that Eug\`{e}ne Reynaud-Bonin, a
former student of this school (ESPT, 1911), has attended to Poincar\'{e}'s
``forgotten lectures'' in 1908 and 1910. His interview was reproduced in \textit{Le Monde}
(see Fig. 8 below).

\begin{figure}[htbp]
\centerline{\includegraphics[width=14.45cm,height=16cm]{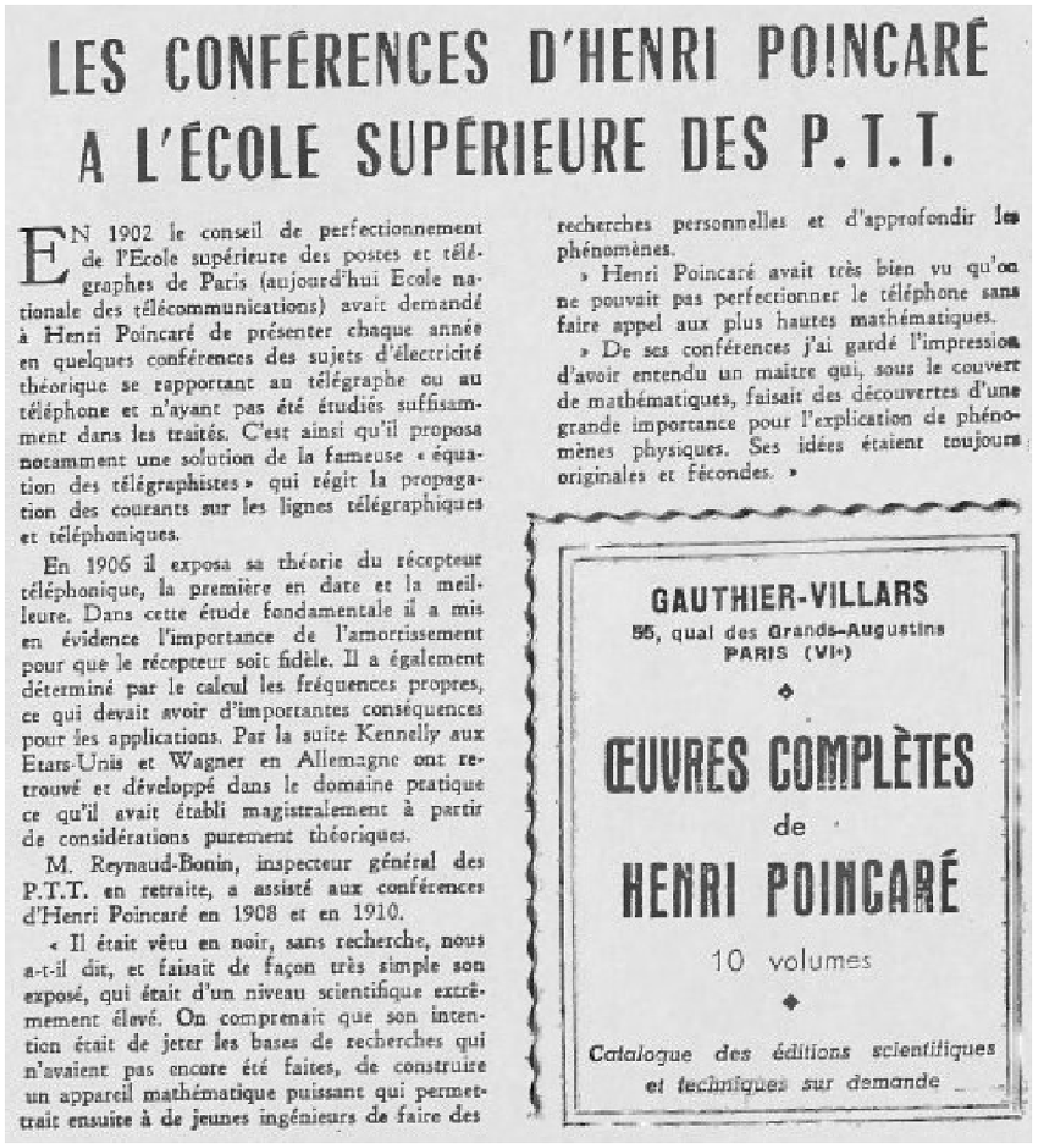}}
\label{fig8}
\caption{\textit{Le Monde}, May 15$^{th}$, 1954}
\end{figure}

Three days later, the physicist Nicolas Minorsky \cite{Minorsky1954} presented a lecture to
civil engineers entitled: \og{}Influence d'Henri Poincar\'{e} sur
l'\'{e}volution moderne de la th\'{e}orie des oscillations non
lin\'{e}aires\fg{} in which he began with these words:

\begin{quote}

\og{}La r\'{e}percussion des travaux d'Henri POINCAR\'{E} s'est fait sentir
dans presque tous les domaines des sciences appliqu\'{e}es, mais c'est
surtout dans la th\'{e}orie des oscillations qu'elle a provoqu\'{e} de tels
changements que cette th\'{e}orie est aujourd'hui passablement
diff\'{e}rente de ce qu'elle soit.\fg{}

\end{quote}

Then, he concluded by this sentence which shows that he didn't have
knowledge of Poincar\'{e}'s ``forgotten lectures'':

\begin{quote}

\og{}Il est difficile de trouver dans l'histoire de la Science un autre
exemple de th\'{e}orie math\'{e}matique d\'{e}velopp\'{e}e sans aucune
relation aux applications {\ldots} qui ait pr\'{e}sent\'{e} une base aussi
parfaite pour l'\'{e}tude des ph\'{e}nom\`{e}nes innombrables qui se sont
r\'{e}v\'{e}l\'{e}s depuis lors, sans qu'il y ait presque rien a changer
\`{a} cette th\'{e}orie un demi-si\`{e}cle plus tard.\fg{}

\end{quote}

\end{document}